\documentstyle[12pt]{article} 
\begin{document} 
\title{Weyl Invariant $p$-brane and D$p$-brane actions} 
\author{J. Antonio Garc\'\i a\thanks{garcia@nuclecu.unam.mx}, Rom\'an Linares\thanks{linares@nuclecu.unam.mx}, and J. David Vergara\thanks{vergara@nuclecu.unam.mx} \\ 
\small\it Departamento de F\'\i sica de Altas Energ\'\i as, \\  
\small \it Instituto de 
Ciencias Nucleares, UNAM, A. Postal 70-543, M\'exico D.F.}   
\date{} 
\maketitle
\begin{abstract}
Conformal invariant new forms of $p$-brane and D$p$-brane actions are
proposed. The field content of these actions
are: an induced metric, gauge fields, an auxiliary metric and an auxiliary
scalar field that implements the Weyl invariance. This scalar field transforms  with a conformal weight that depends on the brane dimension. The proposed actions are Weyl invariant in any dimension and the elimination of the auxiliary metric and the scalar field reproduces the
Nambu-Goto action for $p$-branes and the Born-Infeld action for D$p$-branes. 
As a consequence of the fact that in the $p$-brane case, the action is quadratic in $\partial_0 X$, we develop a complete construction for the associated canonical formalism, solving the problem in previous formulations of conformal $p$-brane actions where the Hamiltonian can not be constructed. We obtain the algebra of constraints  and  identify the generator of the Weyl symmetry. In the construction of the corresponding supersymmetric generalization of this conformal 
$p$-brane action we show that the associated $\kappa$-symmetry is consistent with the conformal invariance. In the D$p$-brane case, the actions are quadratic 
in the gauge fields. These actions can be used to construct new conformal 
couplings in any dimension $p$ to the auxiliary scalar field now promoted as a 
dynamical variable. 
\end{abstract}

\section{Introduction}

The underlying extended objects that allowed recent progress 
in string theory, are 
the $p$-branes \cite{Duff} and Dirichlet branes \cite{Clifford}, or  D$p$-branes.
The $p$-branes are extended structures embedded in a higher dimensional 
space-time from which it inherits an induced metric.
The dynamical properties of such objects are described by the Nambu-Goto
action and its generalizations
for $p$-branes that are now  proportional to the world-volume.
For the case $p=1$
(string theory) the Nambu-Goto action,
given in terms of the area of the string worldsheet can be replaced
 with a classically equivalent
action involving an auxiliary worldsheet metric and local conformal 
symmetry \cite{Howe}. In contrast with the non-polynomial Nambu-Goto action
the new action is quadratic in the derivatives of the 
coordinates. The introduction of the worldsheet metric as an auxiliary field
admits a covariant gauge, 
simplifying the analysis and allowing a covariant quantization \cite{Polyakov}.  
Conformal
invariance or rather Weyl invariance as local
worldvolume symmetry can be implemented for all extended objects, not
just strings ($p=1$). A proposal for the
construction of a Weyl invariant action for any $p$ \cite{DDI} gives
a higher non-polynomial action that prevents the construction
of the associated canonical formalism as well as the study of symmetries and
quantum properties. An attempt to develop the canonical analysis of this type of 
actions was proposed in \cite{barselo}. Here, we present a different solution 
to this problem. We construct a new Weyl invariant action using an auxiliary scalar 
field, for which the standard rules of the canonical analysis can be applied.

On the other hand the D$p$-brane is the ($p+1$) dimensional hypersurface in space-time
where the open strings can end and its dynamics is induced by the open
strings attached on it. 
 They have been crucial in diverse topics like  dualities \cite{Sen}, 
black hole physics \cite{Strominger}, AdS/CFT 
correspondence \cite{Maldacena} and M-theory \cite{Banks}. The action that describes 
the dynamics of these objects is of Born-Infeld (BI) type \cite{Leigh}. 
An action which is quadratic in the gauge fields has been proposed
in \cite{Abou-Hull} --based on the introduction of an auxiliary worldvolume metric
$\gamma_{ij}$-- that is conformal invariant only for the case $p=3$ 
corresponding to D3-branes that have played a central role in recent
studies of D-brane dynamics and string theory in particular for the AdS/CFT
correspondence. The three basic examples of AdS/CFT duality with maximal
supersymmetry are provided by taking the D$p$-branes to be either
M2-branes, D3-branes, or M5-branes. Then the corresponding world volume
theories (in 3, 4, or 6 dimensions) have superconformal symmetry. They are
conjectured to be dual to M theory or type IIB superstring in space time
geometry that is $AdS_4\times S^7, AdS_5\times S^5$, or $AdS_7\times S^4$. 
For example the isometry group of $AdS_5$ is $SO(2,4)$ that is also 
the conformal group in $3+1$ dimensions.

It is possible to construct Weyl invariant actions
associated to D$p$-brane actions for any $p$ by using the same ideas as 
the ones applied to the case of the
$p$-brane. Here we will show that
by introducing a non-dynamical scalar field $\varphi$ with conformal
weight that depends on the dimension we can construct conformal
invariant actions for D$p$-branes. Using the conformal invariance we 
promote this scalar field to a dynamical variable. Possible extensions
of this ideas can be applied  to M-theory
five brane action \cite{Bandos}, and PST action \cite{PST}.

This letter is organized as follows,
in section 2 we introduce the bosonic Weyl invariant $p$-brane action, its
symmetries and  the extended energy momentum tensor. In section 3, 
the canonical analysis of the $p$-brane is work out in detail, obtaining 
the algebra of constraints and identifying the Weyl symmetry generator.
In section 4  we construct a Weyl invariant supersymmetric
version for any $p$ that is quadratic in $\partial_0 X$. The $\kappa$-symmetry
associated to this action is now in addition conformal invariant.
Finally in section 5 we present the corresponding Weyl invariant action for the D$p$-brane, 
and we promote the auxiliary scalar field to a dynamical variable preserving 
the conformal symmetry.   

\section{$p$-brane Actions}

 The propose of this section is to explicitly show
an action for $p$-brane systems, that is by construction Weyl invariant
in any dimension by adding to the auxiliary metric an auxiliary scalar 
field with 
conformal factor that depends
on the worldvolume dimension. The transformation properties for this scalar field
are fixed at the beginning of the analysis to preserve Weyl invariance
and diffeomorphism invariance. The canonical analysis shows a complete consistence
of the constraints algebra  and symmetry generators as in the case
of the non-Weyl invariant $p$-brane action.

The bosonic Nambu-Goto action for a $p$-brane is 
 \begin{equation}\label{A1PB} 
 S[X^{\mu}]=-T \int d^{p+1}\xi \sqrt{-\det g_{ij}},    
\end{equation}
where $T$ is the $p$-brane tension, $\xi^i$, $i=0,1,\dots,p$ 
are the $p+1$ dimensional world volume coordinates and
\begin{equation}
g_{ij}= g_{\mu\nu} \partial_{i}X^{\mu}
\partial_{j}X^{\nu}, 
\end{equation}
is the worldvolume metric induced by the space-time metric $g_{\mu\nu}$.
The non-linear form of this action is inconvenient
for canonical analysis and quantization. For that reason it is useful
to introduce an auxiliary intrinsic
world-volume metric $\gamma_{ij}$ to write down the following quadratic
$p$-brane action
\begin{equation}\label{A2PB} 
S[X^{\mu},\gamma_{ij}] =-\frac{T}{2}\int d^{p+1}\xi \sqrt{-\gamma} 
\left( \gamma^{ij} g_{ij} -(p-1) \right). 
\end{equation} 
Here $\gamma^{ij}$ is the inverse of $\gamma_{ij}$ and $\gamma$ denotes the 
determinant of $\gamma_{ij}$.

This action is invariant under the Weyl transformation 
$\gamma_{ij}\to \exp(2\omega)\gamma_{ij}$ only for the case $p=1$.
The role played by this conformal symmetry was crucial to perform the functional integral of the worldsheet metric reducing the problem to a two dimensional
Liouville theory. The explicit calculation relies on the fact, special to
two dimensions, that it is possible to choice a conformal gauge in such a way
that the integration over $X^\mu$ yield an integral which depends on the
conformal factor through the conformal anomaly. In higher dimensional 
extended objects, conformal
invariance or rather Weyl invariance  
can be implemented through a higher non-polynomial action. This
 Weyl invariant extension of the action for the 
$p$-brane \cite{DDI} is 
\begin{equation}\label{A1PBC} 
S[X^\mu,\gamma_{ij}]=-T\int d^{p+1}\xi \sqrt{-\gamma} 
 \left( \frac{1}{p+1}\gamma^{ij} g_{ij} \right ) ^{(p+1)/2}. 
\end{equation} 
This action has the same or worse drawbacks as the original Nambu-Goto action.
It is non-polynomial and difficult to analyze and quantize. 
The study of this action  by using canonical methods is highly
non-trivial mainly
because the non-linearity inherent in its definition. Indeed, the canonical momenta 
associated to the space-time coordinates,  given by
\begin{equation}
P_\mu = - T \sqrt{-\gamma}  \left( {1\over p+1} \gamma^{ij} g_{ij} \right)^{(p-1)\over 2}
\gamma^{0i} \partial_i X_\mu,
\end{equation} 
does not allow us to find the velocities in terms of the momenta for $p>1$. By this reason
we propose the introduction of a new auxiliary scalar field $\varphi$ 
in addition to
the intrinsic metric $\gamma_{ij}$ with appropriate Weyl weight to preserve
the symmetries of the given action. The action that is now quadratic in 
$\partial  X$ is
\begin{equation}\label{APBCN} 
S[X^{\mu},\gamma^{ij},\varphi] =-\frac{T}{2}\int d^{p+1}\xi \ \sqrt{-\gamma} 
\left (\varphi^{1-\frac{2}{p+1}}\gamma^{ij}g_{ij} -\varphi(p-1) \right). 
\end{equation} 
The above action is invariant under Weyl symmetry for any $p$ 
if the scalar field transforms as
$\varphi\to \exp(-\omega(p+1))\varphi$  while the intrinsic 
worldvolume 
metric $\gamma_{ij}$ transforms as usual with the Weyl weight $2\omega$. 
Notice that in order to preserve
the diffeomorphism invariance of this action the scalar field must also
transform under diffeomorphisms as a scalar. This peculiar transformation
property of the scalar field $\varphi$ should be contrasted with the
corresponding transformation law of the einbein introduced to remove
the square root in the Nambu-Goto action \cite{Schild}.

The action (\ref{APBCN}) is classically 
equivalent to the original Nambu-Goto action upon the elimination of
the auxiliary fields $\varphi$ and $\gamma_{ij}$ by using its own equations of 
motion. It is also classically equivalent to (\ref{A1PBC}) upon the elimination of 
the scalar field $\varphi$.
By fixing the gauge for the Weyl symmetry as 
 $\varphi=1$ we recover the 
action for a $p$-brane given by (\ref{A2PB}).

The 
infinitesimal transformations of the fields that leave this action
invariant up to a boundary term are 
\begin{eqnarray} 
 \delta X^\mu&=&\varepsilon ^i\partial_i X^\mu, \\ 
 \delta \gamma_{ij}&=& \varepsilon^k \partial_k\gamma_{ij} 
+\partial_i\varepsilon^k \gamma_{kj} +\partial_j \varepsilon^k \gamma_{ik}
+2\omega \gamma_{ij},\\
\delta \varphi&=& \varepsilon^i\partial_i \varphi -\omega (p+1)\varphi,  
\label{symm-phi}
\end{eqnarray}  
where $\varepsilon^i$ are $p+1$ infinitesimal arbitrary parameters 
associated with the
diffeomorphism local invariance and $\omega$ is an arbitrary parameter responsible
of the Weyl local symmetry. Notice the transformation property of the scalar 
field (\ref{symm-phi}) under local diffeomorphisms.

It is possible to construct the associated energy 
momentum-tensor for this theory by the prescription
\begin{equation}
T_{ij}\equiv \frac{\varphi^{\frac{2}{p+1}-1}}{\sqrt{-\gamma}}
\left(\frac{\delta S}
{\delta \gamma^{ij}}+\frac{\varphi}{2}\frac{\delta S}{\delta \varphi}\right)
= \frac{1}{p+1}\gamma^{kl}g_{kl}\gamma_{ij}-g_{ij}.
\label{tensor1}
\end{equation}  
This tensor is by definition Weyl invariant and is zero on-shell.
Observe that this is not the standard definition of the energy momentum tensor
because the intrinsic metric and the auxiliary field transforms under the
Weyl symmetry.
 The solution
for $\gamma_{ij}$ is $\gamma_{ij}=\beta g_{ij}$ with $\beta$ an arbitrary 
conformal factor. This tensor has zero trace as a consequence of Weyl 
invariance and it is the same as the one reported in \cite{Lindstrom} up to 
irrelevant factors. 
It could be interesting to work the trace anomaly for
the associated conformal theory in the search of critical dimensions for
these conformal invariant theories. 

We have also worked out the double dimensional reduction 
 \cite{Carmen} for our conformal
action (\ref{APBCN}) in the case $p=2$, to recover the conformal 
invariant string theory. 
Taking into account that our theory
is conformal we can expect some new behavior of this symmetry upon
double dimensional reduction. Nevertheless, we found that the conformal
theory underlying the compactification is not affected by the original
conformal symmetry of the action (\ref{APBCN}).

\section{ Canonical analysis of the Weyl invariant $p$-brane}

 In this section we develop the canonical formulation of the action  
(\ref{APBCN}) in Minkowski space-time and compute the algebra of constraints.  
To construct the associated canonical analysis for the action (\ref{APBCN}) we 
assume that the topology of the  world-volume ${\cal M}^{p+1}$ 
is of the form $\Sigma^p\times \Re$, where $\Sigma^p$ is a $p$ dimensional 
compact manifold. Following the ADM construction we introduce a shift 
vector $N^a$ ($a=1,...,p$) and a lapse function $N$. Using these variables the 
metric of the world-volume $\gamma_{ij}$ can be written as 
\begin{equation}\label{NVFH} 
\gamma_{00}=-N^2 \bar \gamma +\bar \gamma_{ab}N^aN^b, \ \ \ \gamma_{0a}=\bar \gamma_{ab}N^b,
\ \ \ \gamma_{ab}=\bar \gamma_{ab},
\end{equation} 
where $\bar \gamma_{ab}$ is the intrinsic metric of $\Sigma^p$. Substituting 
(\ref{NVFH}) in the Lagrangian action (\ref{APBCN}) we find 
\begin{equation}\label{A2PBNV} 
S=-\frac{T}{2}\int d^{p+1}d\xi  
\left( \frac{\varphi^{\frac{p-1}{p+1}}}{N}\left[-{\dot X}^2 +2 
N^a\dot X^\mu \partial_a X_\mu + \left( N^2\bar \gamma \bar \gamma^{ab} 
-N^aN^b\right)g_{ab} \right] -(p-1)\varphi N\bar \gamma \right). 
\end{equation} 
The associated canonical momenta to the configuration variables 
$(X^{\mu},N,N^a,\bar \gamma^{ab},\varphi)$ are   
\begin{equation}\label{DDMX} 
P_{\mu} =T\frac{\varphi^{\frac{p-1}{p+1}}}{N} 
\left( {\dot X}_{\mu}-N^a\partial_a X_{\mu} \right),  
\end{equation} 
\begin{equation}\label{DDMP} 
\pi_{\ } \approx 0, \hspace{1cm} \pi_a \approx 0, \hspace{1cm} 
\pi_{ab} \approx 0 , \hspace{1cm} \pi_\varphi \approx  0.  
\end{equation} 
The basic Poisson brackets are
\begin{eqnarray} 
\{ X^\mu(\xi),P_\nu(\xi') \} &=& \delta^\mu_\nu \delta^p(\xi-\xi'),  
\hspace{1cm} 
\{ N(\xi),\pi(\xi') \}= \delta^p(\xi-\xi'),  \nonumber \\ 
\{ N^a(\xi),\pi_b(\xi') \}&=& \delta^a_b \delta^p(\xi-\xi'), \hspace{1cm} 
\{ \varphi(\xi),\pi_\varphi(\xi') \}= \delta^p(\xi-\xi'), \\ 
\{ \bar \gamma^{ab}(\xi),\pi_{cd}(\xi') \}&=& 
\frac{1}{2}(\delta^a_c \delta^b_d+\delta^b_c \delta^a_d)\delta^p(\xi-\xi'). 
\nonumber  
\end{eqnarray} 
From the definition of the momenta we obtain $\frac{(p+1)(p+2)}{2}+1$
primary constraints. The total Hamiltonian associated to (\ref{A2PBNV}) is
\begin{eqnarray} 
 H_T&=&\int d^p\xi \left( \frac{N}{2}\left( \frac{1}{T}\varphi^{\frac{1-p}{p+1}} 
P_{\mu}P^{\mu}+T \varphi^{\frac{p-1}{p+1}}\bar \gamma \bar \gamma^{ab}g_{ab}-T\varphi \bar \gamma(p-1) 
\right) +N^a  P_\mu \partial_a X^\mu \right.\nonumber \\ 
& & \left. +\lambda_\varphi \pi_\varphi +\lambda \pi +\lambda^a \pi_a  
+\lambda^{ab} \pi_{ab}\right), 
\end{eqnarray} 
here the $\lambda$'s are the Lagrangian multipliers associated to the primary 
constraints. By using the Dirac method the evolution in time  of these 
constraints generate the following $(p+1)(p+2)/2$ secondary constraints
\begin{eqnarray} 
{\cal H}_{\ \ }&=&\frac{1}{2}\left( \frac{1}{T}P_{\mu}P^{\mu}+
T \varphi^{2-\frac{4}{p+1}}\bar \gamma \bar \gamma^{ab}g_{ab}-
T \varphi^{2-\frac{2}{p+1}}\bar \gamma(p-1)
 \right) \approx 0, \nonumber \\ 
{\cal H}_{a\ }&=&P_\mu \partial_aX^\mu \approx 0, \\ 
\Omega_{ab}&=&T(g_{ab}-\varphi^{\frac{2}{p+1}}\bar \gamma_{ab}) \approx 0. \nonumber  
\end{eqnarray} 
The evolution in time of these secondary constraints does not produce new 
constraints.
To split the constraints according to its first or second class character
we observe that the constraints $\Omega_{ab}$ and $\pi_{ab}$ are second class. 
This are $p(p+1)$ constraints.  Furthermore, from the range of the matrix 
defined by the Poisson brackets between all the constraints, we conclude 
that there are no more second class constraints. 
By a redefinition of  the constraints $\cal H$, ${\cal H}_a$ and 
$\pi_\varphi$ on the constraint surface,
the algebra of the complete set of $2p+3$ first class constraints can be 
closed up to quadratic pieces in second class constraints. 
To that end we propose the following  complete set of constraints 

\noindent {\em First class:}
 \begin{eqnarray} 
\pi &\approx & 0, \ \ \ \pi_a \approx  0, \\ 
{\cal T}_{\varphi} &\equiv & \pi_\varphi+
\frac{2}{p+1}\varphi^{-\frac{2}{p+1}-1}\pi^{ab} 
g_{ab}\approx 0, \\ 
{\cal T} &\equiv & {\cal H} +  
  \frac{2}{T}P_\mu \partial_a(\varphi^{-\frac{2}{p+1}}\pi^{ab}\partial_b 
X^\mu)\approx 0, \\ 
{\cal T}_a &\equiv & {\cal H}_a + 2\partial_a X^\mu  
\partial_b(\varphi^{-\frac{2}{p+1}}\pi^{bc}\partial_c X_\mu)\approx 0. 
\end{eqnarray} 
{\em Second class:} 
\begin{equation} 
 \pi_{ab} \approx  0, \  \ \ 
 \Omega_{ab} \approx  0.  
\end{equation} 
The Poisson bracket between the second class constraints is
\begin{equation}
\{ \Omega_{ab}(\xi), \pi_{cd}(\xi') \} = {1\over 2} (\bar \gamma_{ac}\bar \gamma_{bd}
+ \bar \gamma_{ad} \bar \gamma_{bc}) \varphi^{2\over p+1} \delta^p(\xi -\xi')
\end{equation}

To compute the algebra {\it on} the second class constraint surface we  
introduce the Dirac bracket,
\begin{eqnarray}\label{PDAC} 
\{ F,G \}^* =\{ F,G \}+ {1\over T}\bigg(\int d^p\xi \{ F, \Omega_{ab}(\xi)\}\varphi^{-2\over p+1} \bar \gamma^{ac} 
\bar \gamma^{bd}(\xi) \{ \pi_{cd}(\xi),G\} \nonumber \\ - 
\int d^p\xi \{ F, \pi_{ab}(\xi)\}\varphi^{-2\over p+1} \bar \gamma^{ac} 
\bar \gamma^{bd}(\xi) \{ \Omega_{cd}(\xi),G\}\bigg). 
\end{eqnarray} 
The relevant  Dirac brackets between the  canonical variables are 
\begin{eqnarray} 
\{ X^\mu(\xi),P_\nu(\xi') \}^* &=& \delta^\mu_\nu \delta^p(\xi-\xi'),  
\hspace{1cm} 
\{ N(\xi),\pi(\xi') \}^*= \delta^p(\xi-\xi'),  \nonumber \\ 
\{ N^a(\xi),\pi_b(\xi') \}^*&=& \delta^a_b \delta^p(\xi-\xi'), \hspace{1cm} 
\{ \varphi(\xi),\pi_\varphi(\xi') \}^*= \delta^p(\xi-\xi'), \nonumber \\ 
\{ \bar \gamma^{ab}(\xi),\pi_{cd}(\xi') \}^*&=&0, \\ 
\{ \bar \gamma^{ab}(\xi), P_\mu(\xi') \} ^* &=&   
\varphi^{-\frac{2}{p+1}} \bar \gamma^{ac}\bar \gamma^{bd}  
\left[\partial_d X_\mu \partial_c \delta^p(\xi - \xi')+
\partial_c X_\mu \partial_d \delta^p(\xi - \xi')\right], \nonumber \\ 
\{ \bar \gamma^{ab}(\xi), \pi_\varphi(\xi')\}^*&=& \frac{2}{\varphi(p+1)}\bar \gamma^{ab}
\delta^p(\xi - \xi'). 
\nonumber 
\end{eqnarray} 
Using these Dirac brackets the full Dirac algebra of the densitized 
constraints is 
\begin{eqnarray} 
\{ F[f],\Omega_{ab}[g] \}^*&=&\{ F[f],\pi_{ab}[g] \}^* =0, \nonumber \\ 
\{ {\cal T}_\varphi[f],{\cal T}[g] \}^*&=& 
\{ {\cal T}_\varphi[f],{\cal T}_a[g] \}^* = 0, \nonumber \\ 
\{ {\cal T}[f],{\cal T}[g] \}^*&=& {\cal T}_a[(f\partial_bg-g\partial_bf)
\varphi^{(2-\frac{4}{p+1})}
\bar \gamma 
\bar \gamma^{ab}],  \\ 
\{ {\cal T}[f],{\cal T}_a[g] \}^*&=& {\cal T}[f\partial_a g - g\partial_a f], \nonumber \\ 
\{ {\cal T}_a[f],{\cal T}_b[g] \}^*&=& {\cal T}_b[f\partial_a g] - {\cal T}_a[g\partial_b f]. \nonumber  
\end{eqnarray}
where $F$ is any function of the canonical variables.

From this analysis we can conclude that the
standard diffeomorphism algebra for the $p$-brane  
is reproduced but with different structure functions that are now 
modified to preserve the Weyl symmetry. Compared to the standard non-conformal 
$p$-brane action (\ref{A2PB}) here we have a new  symmetry generator 
$-\omega (p+1) \varphi {\cal T}_\varphi $ for the Weyl local transformation 
through the corresponding Dirac brackets. From constraint analysis we 
obtain the number of physical degrees of freedom in the following way.  
We have  $D+1+(p+1)(p+2)/2$  configuration variables,  
$2p+3$ first class constraints and $p(p+1)$ second class constraints.   
Then the number of physical degrees of freedom per space-time point is,  
$D-(p+1)$.

\section{Conformal Actions for Super $p$-Brane}
In previous sections we show how to build Weyl invariant actions 
for bosonic $p$. In this section we will extend the
construction of the Weyl invariant  $p$-brane action (\ref{APBCN}) to the
supersymmetric Green-Schwarz type actions in curved
space-time.  

Following the construction of the super $p$-brane in \cite{Berg} we add
to the action (\ref{APBCN}) a Wess-Zumino term \footnote{Here $p$ and the space-time 
dimension are restricted by the brane scan \cite{Achu}.} 
\begin{eqnarray}\label{ABCB}
S&&{\hskip-.41cm} [X^\mu,\gamma_{ij}, \varphi]=- {T\over 2} \int d^{p+1}\xi 
\Bigg \{ \sqrt{-\gamma} 
\left ( \varphi^{1-\frac{2}{p+1}}\gamma^{ij}\partial_i X^\mu \partial_j
 X^\nu G_{\mu \nu}
 -\varphi(p-1) \right)   \nonumber \\
& & \hspace{1cm} +  \frac{2}{(p+1)!} 
\varepsilon^{i_1 i_2  \cdots i_{p+1}}\partial_{i_1} X^{\mu_1} 
\partial_{i_2} X^{\mu_2} \cdots
\partial_{i_{p+1}} X^{\mu_{p+1}}
B_{\mu_{p+1} \cdots \mu_2 \mu_1}  \Bigg \},        
\end{eqnarray}
where $G_{\mu\nu}$ is the background metric and $B$ is an antisymmetric 
tensor. This action (\ref{ABCB}) preserves the 
Weyl symmetry under transformations of the intrinsic metric and
auxiliary field proposed in section 2. To obtain the super $p$-brane action, we introduce 
the coordinates $Z^M$ of the curved super-space-time
\begin{equation}
 Z^M=(X^\mu, \theta^\alpha),
\end{equation} 
and the supervielbein $E^A_M(Z)$, where $M=\mu, \alpha$ are
super-space-time indices and $A=a,\alpha$ are indices in the associated 
tangent
space. Defining the supervielbein pull-back
\begin{equation}\label{PBSS}
 E_i^A=\partial_i Z^M E_M^A,
\end{equation}
the action in the superspace can be written as
\begin{eqnarray}\label{PBS}
S[Z^M,\gamma_{ij},\varphi]
&=&- {T\over 2} \int d^{p+1}\xi \left \{ \sqrt{-\gamma} 
\left ( \varphi^{1-\frac{2}{p+1}}\gamma^{ij} E^a_i E^b_j \eta_{ab}
 -\varphi(p-1) \right) \right. \nonumber \\
& & \left. \hspace{2cm} +\frac{2}{(p+1)!} 
\varepsilon^{i_1 \cdots i_{p+1}} E_{i_1}^{A_1} \cdots E_{i_{p+1}}^{A_{p+1}}
B_{A_{p+1} \cdots A_1}  \right \}.        
\end{eqnarray}
This action is invariant under the super world volume diffeomorphisms 
and Weyl transformations 
\begin{eqnarray}
 \delta Z^M &=& \eta^i\partial_i Z^M, \hspace{1cm} 
 \delta \varphi = \eta^i \partial_i \varphi - \omega (p+1)\varphi, \nonumber \\
 \delta \gamma_{ij} &=& \eta^k \partial_k \gamma_{ij} + \partial_i \eta^k \gamma_{jk} 
 + \partial_j \eta^k \gamma_{ik} +2\omega \gamma_{ij},
\end{eqnarray}
where $\eta^i$ are the infinitesimal parameters associated with the diffeomorphism transformation. Furthermore, it is invariant under  supersymmetric
 local $\kappa$ transformations, given by 
\begin{eqnarray}
\nonumber
\delta E^a&=&0, \\
\nonumber
\delta E^\alpha &=&(1+\Gamma)^\alpha_{\ \beta} \kappa ^\beta , \\
\delta(\sqrt{-\gamma} \varphi^{1-\frac{2}{p+1}} \gamma^{ij})&=&
-\frac{4i\zeta}{p!}(1+\Gamma)^\alpha_{\ \beta} 
\kappa^\beta(\Gamma_{a_1\cdots a_p})_{\gamma \alpha} E_n^\gamma 
\varphi^{-\frac{2}{p+1}}\gamma^{ni}
\varepsilon^{j i_1\cdots i_p} E_{i_1}^{a_1}\cdots E_{i_p}^{a_p} \nonumber \\
&+&{2 \over(p+1)!\varphi\sqrt{-\gamma}} (-2i\varphi^{-\frac{2}{p+1}}\gamma^{lm}
 E_l^\alpha E_m^a (\Gamma_a)_{\alpha \beta}+(p+1)\Lambda_\beta)
\kappa^\beta \nonumber \\
\label{TM}
& & \varepsilon^{i_1\cdots i_p i} \varepsilon^{j j_p\cdots j_1}
(E_{i_1}^{a_1} E_{j_1 a_1}\cdots E_{i_p}^{a_p} E_{j_p a_p}+ \nonumber \\
& & \varphi^{\frac{2}{p+1}}E_{i_1}^{a_1} E_{j_1 a_1}\cdots 
E_{i_{p-1}}^{a_{p-1}} E_{j_{p-1} a_{p-1}} \gamma_{i_p j_p} + \cdots +
\varphi^{\frac{2p}{p+1}}\gamma_{i_1 j_1}\cdots \gamma_{i_p j_p}) \nonumber \\
&+& 2\sqrt{-\gamma} \varphi^{1-\frac{2}{p+1}} \gamma^{ij}\delta E^\beta \Lambda_\beta ,
\end{eqnarray}
where  $\kappa^\beta(\xi)$ is a spinor in the space-time and a scalar in 
the worldvolume. Here we have used the notation
\begin{equation}
\delta E^A \equiv \delta Z^M E_M^A,
\end{equation}
where $(\Gamma)^\alpha_{\ \beta}$ and 
$\Lambda_\beta$ are defined in terms of the torsion of the super space
\begin{equation}
 (\Gamma)^\alpha_{\ \beta}={\zeta  \over (p+1)!\varphi\sqrt{-g}}
\varepsilon^{i_1\cdots i_{p+1}} E_{i_1}^{a_1} \cdots 
E_{i_{p+1}}^{a_{p+1}}(\Gamma_{a_1\cdots a_{p+1}})^\alpha_{\ \beta},
\end{equation}
\begin{eqnarray}
 T^{\ \ a}_{\alpha \beta}&=&-2i(\Gamma^a)_{\alpha \beta}, \hspace{1cm} 
 \eta_{c(a}T^{\ \ c}_{\ b)\alpha} = \eta_{ab} \Lambda_\alpha , \nonumber \\
H_{\alpha a_{p+1}\cdots a_1}, &=& 
(p+1)\Lambda_\beta (\Gamma_{a_1\cdots a_{p+1}})^\beta_{\ \alpha}, 
\nonumber \\
H_{\alpha \beta a_p\cdots a_1} &=& 2i\zeta(-1)^{p+1}
(\Gamma_{a_1\cdots a_p})_{\beta \alpha}, \nonumber \\
H_{\alpha \beta \gamma A_1 \cdots A_{p-1}}&=&0,  
\end{eqnarray}
with $\zeta=(-1)^{(p+1)(p-2)/4}$. Notice that the form of this
$\kappa$-transformation is very similar to the previously one found in 
\cite{Berg}, with the remark that in our case the $\kappa$-transformation
(\ref{TM}) preserves the Weyl invariance. The action (\ref{PBS}) is quadratic 
in $\partial_0 X$ allowing also a detailed canonical analysis \cite{w-in-prog}.

\section{Weyl invariant action for D$p$-brane} 
The Born-Infeld-type action for D$p$-brane is
\begin{equation}\label{BI}
S= -T \int d^{p+1} \xi \exp(-\phi) \sqrt{-\det (g_{ij} + 
{\cal F}_{ij})},
\end{equation}
where
\begin{equation}
{\cal F}_{ij} \equiv F_{ij} -B_{ij}, \end{equation}
and $\phi$, $g_{ij}$ and $B_{ij}$ are the pullbacks to the world-volume of 
the background dilaton, metric and NS antisymmetric two-form fields and 
$F_{ij}=\partial_i A_j -\partial_j A_i$ is the field strength of the  $U(1)$ 
world-volume gauge field $A_i$. The action (\ref{BI}) was rewritten in a more geometric way 
in \cite{Abou-Hull,Lindstrom} introducing an 
intrinsic metric $\gamma_{ij}$, obtaining the classically equivalent
form 
\begin{equation}\label{AH}
S = - {T \over 4}  \int d^{p+1}  \xi \exp(-\phi) (-g)^{1/4} (-\gamma)^{1/4} 
\left[ \gamma^{ij}(g_{ij} - g^{kl}{\cal F}_{ik} {\cal F}_{lj}) - (p-3) \right].
\end{equation}
This action is the analog of the Polyakov action 
for the string and has the characteristic that is invariant under Weyl transformations 
\begin{equation}\label{Weylt}
\gamma_{ij} \to \exp(2\omega)\ \gamma_{ij},
\end{equation} 
only for $p=3$. This property can be extended to any dimension using an alternative 
action to (\ref{AH}). The form of the action that is Weyl invariant for any $p$ is
\begin{equation}\label{WIDB}
S = - T \int d^{p+1}  \xi \exp(-\phi) (-g)^{1/4} (-\gamma)^{1/4} 
\left[ {1\over p+1} \gamma^{ij}(g_{ij} - g^{kl}{\cal F}_{ik} 
{\cal F}_{lj})  \right]^{p+1 \over 4}.
\end{equation}
Using the equation of motion for $\gamma_{ij}$, the action (\ref{WIDB}) reduces to the 
Born-Infeld action (\ref{BI}). Also, we can see that for $p=3$ the action (\ref{WIDB}) is 
equal to (\ref{AH}).  However, for any dimensions, we have again the problem 
that this action 
is highly non-linear and then difficult to analyze and quantize. To solve this difficulty we 
introduce, in a similar 
way to the case of the $p$-brane, an auxiliary field $\varphi$ that eliminates the power of 
${p+1 \over 4}$ in the action. The resulting expression is 
\begin{equation}\label{WIWE}
S= -{T \over 4} \int d^{p+1}  \xi \exp(-\phi) (-g)^{1/4} (-\gamma)^{1/4} 
\left[ \gamma^{ij}(g_{ij} - g^{kl}{\cal F}_{ik} 
{\cal F}_{lj})\varphi^{p-3 \over p+1} - (p-3) \varphi  \right].
\end{equation} 
The above action is invariant under the Weyl transformation (\ref{Weylt}) iff 
the auxiliary field $\varphi$ transforms as
\begin{equation}\label{taphi}
\varphi\to \exp\left(-{\omega\over 2}(p+1) \right) \varphi ,
\end{equation} 
and in this way compensates the transformations of the intrinsic metric $\gamma_{ij}$. 
Furthermore, from the equation of motion for $\varphi$ we get
\begin{equation}\label{dbe}
\varphi= \left({1\over p+1} \gamma^{ij}(g_{ij} - g^{kl}{{\cal F}}_{ik} 
{\cal F}_{lj}) \right)^{p+1 \over 4}. 
\end{equation}
Using this expression for $\varphi$ in (\ref{WIWE}) we recover (\ref{WIDB}). This 
shows that both actions are classically equivalent and also Weyl
invariant for any $p$. Nevertheless, the action (\ref{WIWE}) is
quadratic in the 
field strength ${\cal F}_{ij}$ for any dimension whereas (\ref{WIDB}) is  
quadratic only for $p=3$.

For the action (\ref{WIDB}) we define a Weyl invariant energy-momentum tensor
$T^{ij}$ using the expression
\begin{eqnarray}\label{dbt1}
T^{ij} &=& -{1\over T} (-\gamma)^{1\over p+1} {\delta S \over \delta \gamma_{ij}} \nonumber \\
&=& \exp(-\phi) (-g)^{1/4} (-\gamma)^{p+5 \over 4(p+1)}\left( {\gamma\cdot G \over p+1}\right)^{p-3 
\over 4} \left[ \gamma^{ij}\left( {\gamma\cdot G \over p+1}\right) - \gamma^{ki}\gamma^{lj} G_{kl} 
\right], 
\end{eqnarray}
where $ G_{ij} = g_{ij} - g^{kl}{{\cal F}}_{ik} {\cal F}_{lj}$ and $\gamma\cdot G=\gamma^{ij}G_{ij}$ . This tensor is traceless $\gamma_{ij}
T^{ij}=0$ for any dimension as a result of the Weyl invariance. For the action (\ref{WIWE}) we can 
introduce an equivalent energy-momentum tensor that is also
traceless and is given by
\begin{eqnarray}\label{dbt2}
T^{ij} &=& -{1\over T} (-\gamma)^{1\over p+1}\left( {\delta S \over \delta \gamma_{ij}}-
{\varphi \over 4} \gamma^{ij}{\delta  S \over \delta \varphi}\right) \nonumber \\
&=& \exp(-\phi) (-g)^{1/4} (-\gamma)^{p+5 \over 4(p+1)}\varphi^{p-3 
\over p+1} \left[ \gamma^{ij}\left( {\gamma\cdot G \over p+1}\right) - \gamma^{ki}\gamma^{lj} G_{kl} 
\right]. 
\end{eqnarray}
By using the equation of motion (\ref{dbe}), the energy-momentum tensor (\ref{dbt2})
is exactly the same as (\ref{dbt1}).

From this new form of the Weyl invariant action for D$p$-branes (\ref{WIWE}), we can 
construct new conformal couplings of the scalar field with the gauge fields on the brane.
The transformation law for the scalar field under diffeomorphisms
suggest that it can be promoted to a dynamical field playing the role analogous to a  space-time coordinate. Considering that the transformation (\ref{taphi}) is equivalent to a U(1) gauge
transformation with imaginary parameter, a natural way to minimally couple
the scalar field $\varphi$ to the gauge fields $A_i$ is by the introduction
of the Weyl covariant derivative \cite{Deser}, 
\begin{equation}
{\cal D}_i = \partial_i + A_i,
\end{equation} 
and requiring that the vector potential $A_i$ transform under the Weyl 
symmetry as a $U(1)$ gauge connection, {\em i.e.},
\begin{equation}
A_i \to A_i + {p+1 \over 2} \partial_i \omega.
\end{equation} 
As a consequence of this property the covariant derivative turns out to be
\begin{equation}
{\cal D}_i\varphi \to \exp\left( {\omega \over 2}(p+1)\right) 
{\cal D}_i\varphi .
\end{equation}
A Weyl invariant action with the scalar field $\varphi$ promoted to a dynamical variable is
\begin{equation}\label{WIWE1}
S= -{T \over 4} \int d^{p+1}  \xi \exp(-\phi) (-g)^{1/4} (-\gamma)^{1/4} 
\left[ \gamma^{ij}G_{ij}\varphi^{p-3 \over p+1} - (p-3) 
\varphi  \right].
\end{equation}
where the new induced metric is
\begin{equation}\label{newmet}
g_{ij} = g_{\mu\nu} \partial_i X^\mu \partial_j X^\nu + \varphi^{-2}{\cal D}_i\varphi {\cal D}_j\varphi. 
\end{equation}
Notice that in addition to the standard space-time induced metric, the scalar and the gauge fields modify the induced metric on the brane. In this sense these fields induce curvature on the brane. 

By the elimination of the auxiliary metric $\gamma_{ij}$ we recover a  Born-Infeld type 
action (\ref{BI}) with the induced metric given by (\ref{newmet}). Notice that this conformal 
coupling is different from the one proposed in \cite{Deser}, where the induced metric $g_{ij}$ 
does not depends on the scalar field and transform under the conformal symmetry. 
In this model it is also possible to incorporate world-volume dynamical 
gravity in a Weyl invariant way by adding  a fourth derivative Weyl term
or  by using the Einstein term with the compensator $\varphi$
\cite{Deser,Abou}.

In this letter we have constructed Weyl invariant versions of the
$p$-brane and D$p$-brane bosonic actions and the space-time supersymmetric 
extension for the
$p$-brane case by using an auxiliary scalar field. 
One important feature that emerges from the $p$-brane action is that it is 
quadratic in the associated velocities $\dot X$. 
As a consequence of this fact the canonical analysis was work out in detail and it may be possible to obtain some
relevant quantum properties as the  potential existence of a
conformal anomaly associated to the stress tensor (\ref{tensor1}), 
and the analysis of the potential appearance  of critical dimensions.

In our discussion of the space-time supersymmetric case we observe that the 
$\kappa$ symmetry constructed in (\ref{TM}) is now conformal invariant. 
For the spinning membrane case the construction of Weyl invariant actions
using auxiliary scalar fields has been proposed recently in \cite{Castro}.

We have not investigated about if the auxiliary scalar field can be 
used to parameterize the conformal diffeomorphisms that are a subclass of the
remaining symmetries after a covariant gauge has been imposed.

An interesting property of the conformal D$p$-brane action is that it is now 
possible to use the conformal symmetry as a guiding principle to construct
couplings of the auxiliary field with the dynamical fields of the theory. The conformal symmetry induces couplings to the gauge fields and deform the 
induced metric on the brane.

The authors acknowledge partial support from the grants DGAPA-UNAM
IN100397, IN117000 and CONACyT-32431-E.


\begin{thebibliography}{99} 
\bibitem{Duff} M. J. Duff, Lectures given at Theoretical Advanced Study 
Institute in Elementary Particle Physics (TASI 96): Fields, Strings, and 
Duality, Boulder, CO, 2-28 Jun 1996, hep-th/9611203 
\bibitem{Clifford} C. V. Johnson, D-brane primer, Lectures given at 
ICTP, TASI, and BUSSTEPP.  hep-th/0007170 
\bibitem{Howe} L. Brink, P. Di Vecchia, P. Howe. Phys. Lett. {\bf B65} (1976) 
471. \\
 S. Deser, B. Zumino. Phys. Lett. {\bf B65} (1976) 369. 
\bibitem{Polyakov} A. M. Polyakov, Phys. Lett. {\bf B103} (1981) 207,211.
\bibitem{DDI} S. Deser, M. J. Duff and C. J. Isham, Nucl. Phys. {\bf B114}, 
(1976) 29.\\ 
U. Lindstr\"om, Int. J. Mod. Phys. {\bf A3}, 2401 (1988).\\
M.S. Alves, J. Barcelos-Neto, Europhys. Lett. {\bf 7} (1988) 395. 
Erratum-ibid. {\bf 8} (1989) 90. J. A. Nieto, Rev. Mex. Fis. {\bf 34} (1988) 597.

\bibitem{barselo} C. Alvear, R. Amorim and J. Barcelos-Neto, Phys. Lett. {\bf B273}
(1991)415.

\bibitem{Sen} A. Sen, An Introduction to Nonperturbative String Theory,
in Cambridge 1997, Duality and supersymmetric theories 297-413. hep-th/9802051.
\bibitem{Strominger} A. Strominger and C. Vafa Phys. Lett. {\bf B379}
(1996) 99.
\bibitem{Maldacena}J. Maldacena, Adv. Theor. Math. Phys. {\bf 2} (1998) 231.
\bibitem{Banks}T. Banks, W. Fischler, S.H. Shenker and L. Susskind, 
Phys. Rev. {\bf D55} (1997) 5112.
\bibitem{Leigh} R.G. Leigh, Mod. Phys. Lett. {\bf A4} (1989) 2767. 
\bibitem{Abou-Hull} M. Abou Zeid and C. M. Hull, Phys. Lett. {\bf B428} 
(1998) 277.
\bibitem{Bandos} I. Bandos, K. Lechner, A. Nurmagambetov, P. Pasti, 
D. Sorokin and M. Tonin, Phys. Rev. Lett. {\bf 78} (1997) 4332.\\
M. Aganagic, J. Park, C. Popescu and J.H. Schwarz, Nucl. Phys. {\bf B496} (1997)
191.
\bibitem{PST} P. Pasti, D. Sorokin and M. Tonin, Phys. Lett. 
{\bf B398} (1997) 41.
\bibitem{Schild} A. Schild, Phys. Rev. {\bf D16} (1977) 1560.
\bibitem{Lindstrom} U. Lindstr\"om and G. Theodoridis, Phys. Lett. {\bf B208} (1988) 407.
\bibitem{Berg} E. Bergshoeff, E. Sezgin and P. K. Townsend, Phys. Lett.  
{\bf B189} (1987) 75. 
\bibitem{Carmen}
A. Achucarro, P. Kapusta, K.S. Stelle, Phys. Lett. {\bf B232}  (1989) 302.\\ 
 J.A. Nieto, C. Nu\~nez,  Nuovo Cim. {\bf B106} (1991) 1045.  
\bibitem{Achu} A. Achucarro, J. M. Evans, P. K. Townsend and D. L. Wiltshire, 
Phys. Lett. {\bf B198} (1987) 441. 
\bibitem{w-in-prog} J.A. Garc\'\i a, R. Linares and J.D. Vergara, {\it work in 
progress}. 
\bibitem{Deser} S. Deser and G. W. Gibbons, Class. Quantum Grav. {\bf 15} 
(1998) L35.
\bibitem{Abou} M. Abou Zeid, {\it Actions for Curved Branes}, hep-th/0001127.  

\bibitem{Castro} C. Castro, {\it Remarks on the existence of Spinning 
Membrane Actions}, hep-th/0007031. 

\end{thebibliography}
\end{document}